\documentclass[%
 reprint,
 amsmath,amssymb,
 aps,
pra,
floatfix,
]{revtex4-1}

\usepackage{graphicx}
\usepackage{dcolumn}
\usepackage{bm}
\usepackage{amsfonts}
\usepackage{subeqnarray}
\usepackage{mathrsfs}
\usepackage{bbold}
\usepackage{appendix}
\usepackage{tikz}
\usetikzlibrary{arrows}
\usepackage{adjustbox}
\usepackage[caption=false]{subfig}
\begin{document}

\title{Interplay between superconductivity and spin-dependent fields in nanowire-based systems}
\author{J. Baumard$^{1,2}$}
\author{J. Cayssol$^1$}%
\author{A. Buzdin$^{1,3}$}%
\author{F. S. Bergeret$^{2,4}$}

\affiliation{$^1$Univ. Bordeaux, CNRS, LOMA, UMR 5798, F-33405 Talence, France} 
\affiliation{$^2$Donostia International Physics Center (DIPC), Manuel de Lardizabal 4, E-20018 San Sebasti\'an, Spain}
\affiliation{$^3$Sechenov First Moscow State Medical University, Moscow, 119991, Russia}
\affiliation{$^4$Centro de F\'isica de Materiales (CFM-MPC), Centro Mixto CSIC-UPV/EHU, Manuel de Lardizabal 5, E-20018 San Sebasti\'an, Spain}

\date{\today}

\begin{abstract}
The interplay between superconductivity, spin-orbit coupling, and Zeeman or exchange field, is studied theoretically in two different setups: a single wire in which all these fields coexist, and a double wire system in which superconducting pairing and the spin-dependent fields are spatially separated. 
We first explore a magnetoelectric effect, namely the appearance of  anomalous charge supercurrents. We determine the conditions under which such currents are allowed by symmetry and express them in terms of the SU(2) electric and magnetic fields. In leading order in the strength of the fields we find that in the single wire setup such currents may appear only when the Zeeman field has both, a longitudinal and  transverse component with respect to the spin-orbit field. In contrast, in the two wire setup a parallel component to the SOC can generate the anomalous current, which is allowed by symmetry. We confirm these findings by calculating explicitly the current in both setups together with the self-consistent superconducting order parameter. The latter shows in the ground-state a spatial modulation of the phase that leads to currents that compensate the anomalous current, such that in both cases the ground state corresponds to a total zero-current state.  However, in the two wire setup this zero-current state consists of two finite currents flowing in each of the wires in opposite direction.  
\end{abstract}
\maketitle

\section{\label{sec:introduction}Introduction}
Superconductivity in low dimensional systems with 
spin-orbit coupling (SOC) and a Zeeman field has been intensively studied in the context of Majorana zero-energy modes \cite{lutchyn_majorana_2010,oreg_helical_2010,alicea_new_2012,beenakker_search_2013,zha_majorana_2015,reeg_transport_2017}. On the other hand the interplay between superconductivity and spin-dependent fields is known to lead to striking phenomena as long-range triplet correlations \cite{bergeret_singlet-triplet_2013,bergeret_spin-orbit_2014}, critical field enhancement \cite{samokhin_magnetic_2004,barzykin_inhomogeneous_2002} and  magnetoelectric effects stemming from the coupling between charge and spin degrees of freedom  \cite{buzdin_direct_2008,ojanen_magnetoelectric_2012,jacobsen_critical_2015,konschelle_theory_2015,nesterov_anomalous_2016,mironov_spontaneous_2017,bobkova_gauge_2017}. 

Magnetoelectric effects in the superconducting state are reminiscent of the widely studied spin Hall effect (SHE) and Edelstein effect in normal conductors \cite{dyakonov_possibility_1971, dyakonov_current-induced_1971,aronov_nuclear_1989,edelstein_spin_1990}. The SHE consists in the generation of a transverse spin current in response to a longitudinal charge current, while a spin density is induced by a charge current in the Edelstein effect. In particular, the Edelstein effect has been theoretically discussed in 2D systems with Rashba SO interaction, and due to possible applications in spintronics, it has been  investigated over the past years \cite{kato_current-induced_2004,silov_current-induced_2004,rojas_sanchez_spin--charge_2013,shao_strong_2016,song_observation_2017}. The inverse Edelstein effect, also known as spin-galvanic effect, describes the generation of an electric current via a spin density \cite{edelstein_magnetoelectric_1995,yip_two-dimensional_2002,rojas_sanchez_spin--charge_2013}. 

These effects also occur in the superconducting state \cite{edelstein_magnetoelectric_1995,yip_two-dimensional_2002,buzdin_direct_2008,bergeret_theory_2015,konschelle_theory_2015, baumard_generation_2019,rabinovich_magnetoelectric_2019}.  For temperatures close to the superconducting critical temperature, the magnetoelectric effect can be studied at the level of the Ginzburg-Landau free energy. Specifically the coupling between spin and charge degrees of freedom is described by an additional term, $F_\text{L}$, the so-called Lifshitz invariant \cite{edelstein_ginzburg_1996, agterberg_magnetoelectric_2012}. For the particular case of a 2D system with a Rashba SOC, this term has the form  $F_\text{L} \sim \varepsilon(\alpha)\,\vec{n}\cdot\left(\vec{h}\times\vec{\nabla}\varphi\right)$, where $\varepsilon(\alpha)$ is a function of the Rashba SOC parameter $\alpha$, $\vec{h}$ is the exchange or Zeeman field, $\varphi$ is the phase of the superconducting order parameter, and $\vec{n}$ is a unit vector perpendicular to the plane. The above expression for the Lifshitz invariant can be generalized for arbitrary linear-in-momentum SOC and dimension \cite{agterberg_microscopic_2014,bergeret_theory_2015}. 

In general, the Lifshitz invariant can be written as 
\begin{equation}
    F_L\sim \vec{T}\cdot \vec{\nabla} \varphi\;,
    \label{eq:F_L_intro}
\end{equation}
where $\vec{T}$ is a polar vector which is odd under time reversal, and SU(2) gauge invariant. This vector $\vec{T}$ can be expressed in terms of the so-called SU(2) magnetic and electric fields. For example in the case of a homogeneous 2D system the current induced via magnetoelectric effect is, in lowest order in the SOC and Zeeman fields, proportional to the vector product between the electric and magnetic SU(2) fields \cite{bergeret_theory_2015}. 
In a one-dimensional system, the SU(2) magnetic field is zero and therefore the situation has to be somehow different. Indeed it was shown in Refs. \cite{ojanen_magnetoelectric_2012, nesterov_anomalous_2016} that the spin-galvanic effect in a 1D wire with Rashba SOC only appears if the Zeeman field has finite components, both parallel and orthogonal to the SOC. This result suggests that there has to be a contribution to the Lifshitz invariant which only contains terms proportional to the SU(2) electric field. 

In the present work, we investigate the charge current originating from the interplay between superconductivity, SOC and a Zeeman field in different systems of wires. Specifically, we determine the Lifshitz invariant in terms of the SU(2) fields, and focus on the charge current to the leading order in the SOC and Zeeman field. We use this general method to study two different setups: a single wire with superconducting correlations, and a system where the superconducting and magnetic correlations and fields are spatially separated into two different wires which are coupled via a hopping term. In leading order in the fields, the single wire system displays a spontaneous current only if both transverse and longitudinal components (with respect to the spin-orbit field) of the Zeeman field are finite. On the other hand, in the two wire system, the parallel component is sufficient to generate a current. This result can be  explained by the existence of an effective SU(2) magnetic field.  

This article is organized as follows: in Sec. \ref{sec:symmetry}, we investigate the appearance of the charge current within the SU(2)-covariant formalism, by constructing a general Lifshitz invariant $F_\text{L}$ in lowest order in the SU(2) electric and magnetic fields. In Sec. \ref{sec:one wire}, we study the single wire system in the framework of Green's function formalism and  compute the current explicitly in the limit of weak spin-orbit coupling and Zeeman field. By allowing a superconducting order parameter with an inhomogeneous phase we demonstrate that the ground-state of the system corresponds to a zero-current state due to the cancellation of the spin-galvanic current by the one generated by the inhomogeneous superconducting phase. Sec. \ref{sec:two wires} is devoted to the double wire system. We derive the self-consistency equation and the expression of the current, and show that the zero-current state carried in the ground state corresponds to two opposite currents flowing in each of the wires. We finally present our conclusions in section \ref{sec:conclusion}.

\section{\label{sec:symmetry}The spin-galvanic effect in superconducting wires from a SU(2)-covariant perspective}

In this section, we first provide a brief  review of  the SU(2) covariant formalism for 2D systems with Rashba SOC and an exchange field. We then apply this formalism to different systems of wires to understand the magnetoelectric effects from a symmetry consideration. 

\subsection{\label{subsec:reviewSU2}Short review of the SU(2) formalism}

To understand the magnetoelectric effect from a general perspective, we briefly introduce the SU(2)-covariant formalism \cite{tokatly_equilibrium_2008,gorini_non-abelian_2010} to describe systems with linear-in-momentum SOC and a quadratic electronic band. The starting point is the general Hamiltonian describing a system with superconducting correlations:
\begin{equation}
    \mathcal{H} = \int\mathrm{d}\vec{r} \left(\Psi^\dagger\,\hat{h}_0\,\Psi + V\,\psi^\dagger_\uparrow\,\psi^\dagger_\downarrow\,\psi_\downarrow\,\psi_\uparrow\right) \;,
    \label{eq:H general}
\end{equation}
where the spinor $\Psi = \left(\psi_\uparrow,\,\psi_\downarrow\right)^T$ contains the annihilation operators $\psi_{\uparrow,\downarrow}(\vec{r})$ for spin up and down electrons at position $\vec{r}$ and $V = V\left(\vec{r}\right)$ is the superconducting potential.  
In the normal part of the Hamiltonian, $\hat{h}_0$, the linear-in-momentum SOC and Zeeman field enter as a SU(2) gauge potential \cite{tokatly_equilibrium_2008,gorini_non-abelian_2010}  
\begin{equation}
    \hat{h}_0 = \frac{\left(p_i - \mathcal{A}_i\right)^2}{2\,m} - \mu + \mathcal{A}_0 \;,
    \label{eq:h0}
\end{equation}
where $\mu$ is the chemical potential. Namely $\mathcal{A}_0 = \displaystyle\frac{1}{2}\,\mathcal{A}_0^a\,\sigma^a$ is the SU(2) scalar potential which describes either a Zeeman magnetic field in a normal metal or an intrinsic exchange field in a ferromagnetic metal, whereas the vector potential $\mathcal{A}_i = \displaystyle\frac{1}{2}\,\mathcal{A}_i^a\,\sigma^a$ represents spin-orbit interaction, where $\sigma^a$ are the Pauli matrices. 
Lower (upper) indices $i = x,\,y,\,z$ label space (spin) variables
and sum over repeated indices is implied. 

In analogy to usual electrodynamics, the Hamiltonian Eq. (\ref{eq:h0}) is written  in terms of a SU(2) four-potential $\mathcal{A}_\mu$, where $\mu$ labels space ($\mu = x,\, y,\, z$) via spin-orbit interaction and time ($\mu = 0$) via the Zeeman field. This Hamiltonian is invariant under the gauge transformations $\Psi \rightarrow \mathcal{U}\,\Psi$ and $\mathcal{A}_\mu \rightarrow \mathcal{U}\,\mathcal{A}_\mu\,\mathcal{U}^{-1} - \text{i}\left(\partial_\mu \mathcal{U}\right)\mathcal{U}^{-1}$, where $\mathcal{U}$ is a SU(2) rotation  matrix \cite{tokatly_equilibrium_2008}. Following the analogy to electrodynamics one  defines the SU(2) field strength tensor $\mathcal{F}_{\mu\nu}$ as:
\begin{equation}
    \mathcal{F}_{\mu\nu} = \partial_\mu\mathcal{A}_\nu - \partial_\nu\mathcal{A}_\mu - \text{i}\left[\mathcal{A}_\mu,\,\mathcal{A}_\nu\right] \;.
    \label{eq:F_mu_nu}
\end{equation}
The SU(2)  electric and magnetic fields are then defined as $E^a_k={\cal F}_{0k}^a$, and $B^a_k=\epsilon_{kij}{\cal F}_{ij}^a$, respectively, $\epsilon_{ijk}$ being the Levi-Civita symbol. 

We consider now a superconductor at temperature close to its critical temperature $T_c$. Equilibrium properties can be described within the Ginzburg-Landau theory. As mentioned in the introduction magnetoelectric effect is related to a Lifshitz invariant in the free energy with the form of Eq. (\ref{eq:F_L_intro}). The anomalous current induced in a system is then proportional to the polar vector $\vec{T}$. The latter can be constructed using SU(2) gauge symmetry arguments only. In Ref. \cite{bergeret_theory_2015}, the Lifshitz invariant in the lowest order in the exchange field was identified, and the anomalous current was found to be proportional to the following cross product of the SU(2) electric and magnetic fields:  
\begin{equation}
    j_i \sim \mathcal{F}_{0k}^a\,\mathcal{F}_{ki}^a = E^a \times B^a \;.
    \label{eq:j_i general}
\end{equation}

To illustrate the  result of expression Eq. (\ref{eq:j_i general}), let us  consider a two-dimensional system in the $x-y$ plane with homogeneous and time-independent Zeeman and Rashba spin-orbit fields. Specifically, the SOC is described by the Hamiltonian $\mathcal{H}_{so} = \alpha\left(p_y\,\sigma^x - p_x\,\sigma^y\right)$ leading to a two component vector potential: $\mathcal{A}_x^y = 2\,m\,\alpha = -\mathcal{A}_y^x$. In this case, the electric field has two components $\mathcal{F}_{0x} = m\,\alpha\left(\mathcal{A}_0^x\,\sigma^z - \mathcal{A}_0^z\,\sigma^x\right)/2$ and $\mathcal{F}_{0y} = m\,\alpha\left(\mathcal{A}_0^y\,\sigma^z - \mathcal{A}_0^z\,\sigma^y\right)/2$, whereas the magnetic field is given by $\mathcal{F}_{xy} = m^2\,\alpha^2\,\sigma^z = -\mathcal{F}_{yx}$. This leads to an in-plane current :
\begin{equation}
{j}_{x,y} \sim \pm \,m^3\,\alpha^3\,\mathcal{A}_0^{y,x} \, ,
\end{equation}
when the Zeeman field is applied in $y$ or $x$ direction respectively,
in agreement with the Edelstein result, Ref. \cite{edelstein_ginzburg_1996}.

The situation is rather different in a 1D system, for which the SU(2) magnetic field $B^a$ is zero and therefore, contribution from  Eq. (\ref{eq:j_i general})  vanishes.  It is however known that in wires with a 1D SOC such current can be  finite \cite{ojanen_magnetoelectric_2012, nesterov_anomalous_2016}, namely, when at least two components of the magnetic field, one longitudinal and one  transverse to the spin-orbit field, are finite. 
This result can also be understood from the above SU(2) arguments if we seek for an invariant of higher order in the exchange field:  clearly in a pure 1D system only the electric field is finite and the next leading order contribution to the current can be constructed as a product of derivatives of the electric field \footnote{I. V. Tokatly, private correspondence}: 
\begin{equation}
    j_i \sim \left(\tilde{\nabla}_k\mathcal{F}_{0k}\right)^a\,\left(\tilde{\nabla}_0\mathcal{F}_{0i}\right)^a \;,
    \label{eq:j_i 1D}
\end{equation}
where $\tilde{\nabla}_k$ denotes the covariant derivatives defined as  $\tilde{\nabla}_\mu = \partial_\mu\cdot - \text{i}\left[\mathcal{A}_\mu,\,\cdot\right]$.

In the next two subsections we discuss two different situations in which either one of the contributions Eqs. (\ref{eq:j_i general} - \ref{eq:j_i 1D}) contributes to the anomalous current. Before that it is important to emphasize that in systems in which superconductivity and the spin-dependent fields spatially coexist, one should compute the superconducting order parameter self-consistently allowing for a spatially dependent phase. Generally, the  ground-state corresponds to a zero-current state \cite{dimitrova_theory_2007} and therefore the anomalous current obtained above has to be compensated by the current induced by the phase gradient of the condensate wave function. In subsequent sections we discuss this in detail.

\subsection{\label{subsec:symmetry 1w} Single superconducting wire}

As mentioned above,   in a system with  a pure 1D SOC  only the SU(2) electric field is finite and hence the anomalous current (in lowest order in the exchange field) is given by Eq. (\ref{eq:j_i 1D}). 
Specifically the electric field for a 1D system  with  time-independent  fields is given by 
\begin{equation}
    \mathcal{F}_{0x}=-\partial_x\mathcal{A}_0 - \text{i}\left[\mathcal{A}_0,\,\mathcal{A}_x\right] \;.
    \label{eq:E-field_1D}
\end{equation}
The second term is the field induced for example by homogeneous exchange field and SOC. The first term is finite for an inhomogeneous exchange field induced for example by a spin texture. 

We first consider a homogeneous situation, for which the exchange field has an arbitrary direction and the SOC is, without loss of generality, $\mathcal{H}_\text{so} = \alpha\,p\,\sigma^z$, such that the vector potential has only one finite component $\mathcal{A}_x^z = -2\,m\,\alpha$. In this case $\mathcal{F}_{0x}= - \text{i}\,m\,\alpha\,\mathcal{A}_0^a \left[\sigma^a,\,\sigma^z\right]/2$ and hence the anomalous current obtained from Eq. (\ref{eq:j_i 1D}) reads:
\begin{equation}
    j_x \sim -m^3\,\alpha^3\left[\left(\mathcal{A}_0^x\right)^2 + \left(\mathcal{A}_0^y\right)^2\right] \mathcal{A}_0^z \;.
    \label{eq:jx_soc_1D}
\end{equation}
This result coincides with the one obtained in Refs.\cite{ojanen_magnetoelectric_2012,nesterov_anomalous_2016} in leading order in the exchange field. 

A gauge equivalent situation to the previous one is  a system without SOC but with an inhomogeneous exchange field that  varies in  $x$-direction \cite{bergeret_spin-orbit_2014}. 
In such a case  
the electric field is finite due to the first term in Eq. (\ref{eq:E-field_1D}). 
As a specific example we consider an exchange field with the following spatial dependence: $\vec{h} = (h_0\cos\left(Q\,x\right),h_0\sin\left(Q\,x\right),h_z)$. The corresponding SU(2) scalar potential has then three components: $\mathcal{A}_0^x = 2\,h_0\,\cos\left(Q\,x\right)$, $\mathcal{A}_0^y = 2\,h_0\,\sin\left(Q\,x\right)$ and $\mathcal{A}_0^z = 2\,h_z$, and according to Eq. (\ref{eq:E-field_1D}), $\mathcal{F}_{0x}= 2\,Q\,h_0\,\left[\sin\left(Q\,x\right)\sigma^x - \cos\left(Q\,x\right)\sigma^y\right]$. 
From Eq. (\ref{eq:j_i 1D}) we obtain:
\begin{equation}
    j_x \sim Q^3\,h_0^2\,h_z\;.
    \label{eq:jx_helical_1D}
\end{equation}
Expressions (\ref{eq:jx_soc_1D}) and (\ref{eq:jx_helical_1D}) coincides after identifying the following equivalences:  $m\,\alpha\leftrightarrow Q$  and  $h_0^2=\left(\mathcal{A}_0^x\right)^2 + \left(\mathcal{A}_0^y\right)^2$.

\subsection{\label{subsec:symmetry 2w}Double wire system}
\begin{figure*}
\begin{center}
\subfloat[\label{fig:one_wire}]{\adjustbox{raise=-11.35pc}{\includegraphics[scale=0.40,]{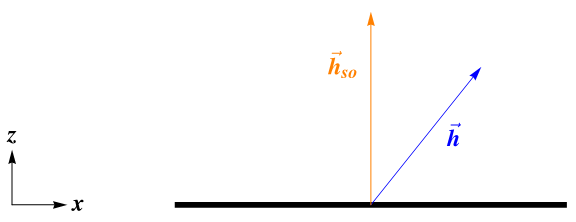}}}
\hspace{0.5cm}
\subfloat[\label{fig:two_wires}]{\includegraphics[scale=0.4]{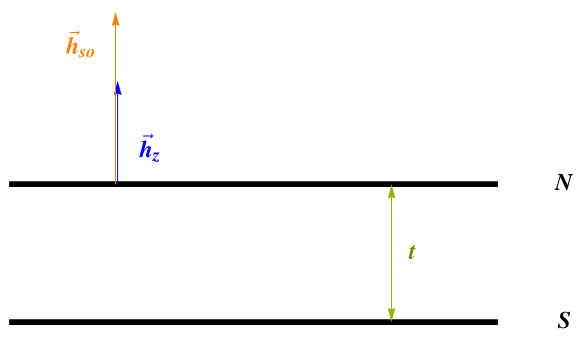}}
\end{center}
\caption{\protect\subref{fig:one_wire} The first setup is made of an infinite superconducting wire with SOC and a Zeeman field with two components, one longitudinal and one transverse to the spin-orbit field. \protect\subref{fig:two_wires}  In the second system, superconducting and magnetic correlations and fields are spatially separated into two different wires, coupled via a hopping term. We consider the case where the spin-orbit and Zeeman fields are parallel. \label{fig:setups}}
\end{figure*}

We now  consider a two wire setup, as the one  sketched in Fig. \ref{fig:two_wires}. One of the wires is a superconductor with no spin-dependent fields, whereas the second wire is in the normal state with a finite SOC and exchange or Zeeman field. Both wires are tunnel coupled. The tunneling is described by a hopping parameter $t$ which is assumed to be spin independent. This coupling, on the one hand, induces superconducting correlations in the normal wire, and on the other hand an effective spin-dependent field is induced in the superconducting wire.    

In a pure 1D system, when all interactions take place in one and the same wire, if the SOC and exchange field vectors are parallel, all SU(2) fields are zero and hence no magnetoelectric effects can take place.  
In the two wire system the situation is rather different, because strictly speaking the system is not pure 1D (indeed there is an additional internal degree of freedom which is the wire index) and the coupling between the wires prevents such a pure gauge situation. 

Namely, the fields depend not only on the spatial coordinate along the wires, but also on the one perpendicular to them denoted as $z$ in Fig. \ref{fig:two_wires}. Then the system behaves as a 2D system. If we assume that SOC and exchange field are parallel, with the only non-zero components of the potentials being $\mathcal{A}_x^z$ and $\mathcal{A}_0^z$, then according to Eq. (\ref{eq:F_mu_nu}) the SU(2) fields are finite and determined by $\mathcal{F}_{0z}^z = -\partial_z\mathcal{A}_0^z$ for the electric field and $\mathcal{F}_{zx}^z = \partial_z\mathcal{A}_x^z$ for the magnetic field. Therefore, from the above symmetry arguments, specifically from Eq. (\ref{eq:j_i general}),  one expects a finite anomalous current, linear in both, the SOC and the exchange field,  even if they are parallel. 

The above analysis only explains whether  the anomalous currents are allowed by symmetry or not. In the next sections we compute explicitly the currents and also the ground state of the setups of Fig. \ref{fig:setups}. 

\section{\label{sec:one wire}Superconducting wire with Zeeman and spin-orbit interactions}
In this section, we focus on a single  superconducting wire in the presence of Rashba SOC and a Zeeman field. Similar setup has been studied in Refs. \cite{ojanen_magnetoelectric_2012, nesterov_anomalous_2016,baumard_non-uniform_2019}. {  We first consider that superconductivity is proximity-induced in the wire, and thus that the superconducting order parameter is constant. In this case, the charge current corresponds to the anomalous current. We compute this current explicitly and compare the result from the one obtained by the above symmetry arguments.  Then, we explore the possibility of a wire with intrinsic superconductivity, and allow for a superconducting order parameter with an inhomogeneous phase. We show explicitly that, within our approximation,  in the ground-state, the anomalous current is exactly compensated by the current induced by the order parameter phase gradient.
Although it is well-known that quantum fluctuations would kill superconductivity in such low-dimensional systems, it has been reported in quasi-1D materials like some organic superconductors \cite{buzdin_organic_1984}. Such compounds consist of weakly coupled 1D chains. If the hopping parameter $\tau$ describing inter-chain coupling is much smaller that $T_c$, then the system can be described by a strictly 1D chain.  It has been established that the mean-field treatment is justified if $\tau \geq T_c^2/\mu$ \cite{tsuzuki_long-range_1972}. So, for $T_c^2/\mu \leq \tau \leq T_c$, the critical fluctuations of the superconducting order parameter are effectively suppressed and the system can be treated as a purely superconducting 1D wire. But even in more conventional systems involving superconducting thin wire, not necessary strictly 1D,  our results provide a  qualitative description of possible magnetoelectric effects.}

{\subsection{\label{subsec:anomalous_1w}Anomalous current}}
We assume that the Zeeman field has  two components: one normal and one  parallel to the SOC field (Fig. \ref{fig:one_wire}). 
The Hamiltonian describing  an infinite  wire  along the $x$-direction reads:
\begin{equation}
    \mathcal{H} = \int \mathrm{d}p\, \Psi^\dagger\left(\hat{h}_\text{N} + \hat{h}_\text{BCS}\right)\Psi \;,
    \label{eq:Hamiltonian 1w}
\end{equation}
where $\Psi = \left(\psi_\uparrow(p),\,\psi_\downarrow(p),\,-\psi^\dagger_\downarrow(-p),\,\psi^\dagger_\uparrow(-p)\right)^T$ is the spinor containing the annihilation and creation operators in Nambu spin basis. The Hamiltonian operator $\hat{h}_\text{N}$ describes the system in the normal state. We assume that the Zeeman field is  in the $x-z$ plane, $\vec{h} = \left(h_x,\,0,\,h_z\right)$, and that the wire lies  on a substrate parallel to this plane, such that the SOC is along the $z$-direction:
\begin{equation}
    \hat{h}_\text{N} = \left(\xi + \alpha\,p\,\sigma_z\right)\tau_z + \vec{h}\cdot\vec{\sigma} \;.
    \label{eq:h_N 1w}
\end{equation}
Here $\displaystyle\xi = \frac{p^2}{2\,m} - \mu$ is the quasiparticle energy, $\mu$ is the chemical potential, $\alpha$ is the spin-orbit coupling constant and $\sigma_{x,y,z}$, $\tau_{x,y,z}$ are the Pauli matrices acting respectively in spin and Nambu space. The BCS Hamiltonian operator $\hat{h}_\text{BCS}$ is expressed as:
\begin{equation}
    \hat{h}_\text{BCS} = - \Delta_0\,\tau_x \;,
    \label{eq:h_BCS 1w}
\end{equation}
where $\Delta_0$ is the superconducting order parameter, first assumed to be a constant. 

To compute the anomalous charge current we use the Green's function formalism \cite{abrikosov_methods_1965}. Close to the normal-superconducting phase transition, $\Delta_0 \ll T$, one can expand the Green's function $G$ in series of $\Delta_0$. In second order in $\Delta_0$, $G$ reads:
\begin{equation}
    G \approx G_\text{N} + G_\text{N}\,\hat{h}_\text{BCS}\,G_\text{N} + G_\text{N}\,\hat{h}_\text{BCS}\,G_\text{N}\,\hat{h}_\text{BCS}\,G_\text{N} \;,
    \label{eq:G 1w}
\end{equation}
where  $G_\text{N}$ is the Green's function  in the normal state  obtained from the Dyson equation $\left(\text{i}\,\omega_n - \hat{h}_\text{N}\right) G_\text{N} = \mathbb{1}$, where $\omega_n = \pi\,T\left(2\,n + 1\right)$ are the Matsubara frequencies at temperature $T$, $n$ being an integer. In Nambu space, $G_\text{N}$ reads
\begin{equation}
    G_\text{N} = \begin{pmatrix}
    G_- && 0\\
    0 && G_+
    \end{pmatrix} \;,
    \label{eq:GN_1w}
\end{equation}
where the matrix $G_\lambda$ ($\lambda = \pm$) is defined in the spin basis: 
\begin{equation}
    G_\lambda = \frac{\text{i}\,\omega_n + \lambda\,\xi + \vec{h}\cdot\vec{\sigma} - \lambda\,\alpha\,p\,\sigma_z}{\left(\text{i}\,\omega_n + \lambda\,\xi\right)^2 - \left(h_z - \lambda\,\alpha\,p\right)^2 - h_x^2} \;.
    \label{eq:G_lambda}
\end{equation}

The charge current  is proportional to the expectation value of the velocity operator $\hat{v} = \displaystyle\frac{\partial \hat{h}_\text{N}}{\partial p}$, 
such that in terms of the Green's function $G$, the charge current reads:
\begin{equation}
    j_\text{an} = -\frac{e\,T_c}{2}\,\sum_{\omega_n}\int_{-\infty}^{+\infty}\frac{\mathrm{d}p}{2\,\pi}\left[\frac{p}{m}\,\text{Tr}\,G + \alpha\,\text{Tr}\left(G\,\sigma_z\right)\right] \;,
    \label{eq:j 1w}
\end{equation}
where $T_c$ is the critical temperature. The second term in Eq. (\ref{eq:j 1w}) stems from the so-called anomalous velocity due to the SOC. 

We are interested in determining the current in the leading order of the spin-dependent fields.  Therefore in what follows   we assume that both, the  Zeeman field and SOC, are small such that   $h_x,\, h_z \ll T_c$ and $\alpha\,p_\text{F} \ll T_c$ (where $p_\text{F}$ is the Fermi momentum) and expand the expressions for the Green's functions $G$ from Eq. (\ref{eq:G 1w}--\ref{eq:G_lambda}) in series of $h_x$, $h_z$ and $\alpha\,p$. 
We then substitute this expansion into the current expression, Eq. (\ref{eq:j 1w}). The integral over momentum has to be done with certain care. First, only terms even in $p$ contribute to the integral that can be written for only positive values of $p$. In turn, this integral can be  transformed into an integral over the quasiparticle energy:
\begin{equation}
\int_{-\infty}^{+\infty} \mathrm{d}p =  2\,\int_{-\mu}^{+\infty} N\left(\xi\right) \mathrm{d}\xi  \;,
\label{eq:change_int}
\end{equation}
where $N\left(\xi\right) = \displaystyle\sqrt{\frac{m}{2\left(\xi + \mu\right)}}$ is the density of states at energy $\xi$. We assume that the chemical potential is the largest energy involved in the problem : $\mu \gg T_c,\, E_\text{so}$, where $E_{so} = m\,\alpha^2/2$, and set the lower integration limit to $-\infty$. Because in the expansion of the Green's function $G$ one keeps terms up to order $\alpha^3\,p^4$ where $p^2 = 2\,m\left(\xi + \mu\right)$, in the expansion of $N\left(\xi\right)$, Eq. ($\ref{eq:change_int}$), one needs to keep also terms up to order $(\xi/\mu)^2$. The integral in Eq. ($\ref{eq:change_int}$) can then be written as:
\begin{equation}
\int_{-\infty}^{+\infty} \mathrm{d}p \rightarrow \frac{2}{v_\text{F}} \int_{-\infty}^{+\infty} \left(1 - \frac{\xi}{2\,\mu} + \frac{3\,\xi^2}{8\,\mu^2}\right) \mathrm{d}\xi  \;,
\label{eq:int_xi}
\end{equation}
where $v_\text{F} = \displaystyle\sqrt{\frac{2\,m}{\mu}}$. After this transformation the  $\xi$ integration of Eq. (\ref{eq:j 1w}) can be performed straightforwardly leading to: 
\begin{equation}
     j_\text{an} = 28\,e\,T_c\,\Delta_0^2\,m^2\,v_\text{F}\,\alpha^3\,h_x^2\,h_z\,\Gamma_7 \;,
     \label{eq:j0 1w}
\end{equation}
where $\Gamma_s = \displaystyle\sum_{\omega_n > 0}\frac{1}{\omega_n^s}$. 
This result implies  the presence of a finite charge current along the wire, in agreement with our symmetry arguments in Sec. \ref{subsec:symmetry 1w} based on  the SU(2) covariant formalism.  Indeed, in  leading order in SOC and Zeeman field, the anomalous current is proportional to $\alpha^3\,h_x^2\,h_z$. 

{ \subsection{\label{subsec:inhomogeneous_1w}Charge current for an inhomogeneous superconductor}}
Now we allow for an inhomogeneous superconducting state with an  order parameter $\Delta (x) = \Delta_0\,e^{\text{i}\,q\,x}$. {  We assume that $q\,v_\text{F} \ll T_c$, {\it i.e.} when the system is near the emergence of the modulated phase.
In this case   the total current can be written as :}
\begin{equation}
    j = j_\text{an} + j_q \;,
    \label{eq:transformed j 1w}
\end{equation}
{where $j_\text{an}$ is the anomalous current, obtained for $q = 0$, and }
\begin{equation}
    j_q = - 2\,e\,T_c\,\Delta_0^2\,v_\text{F}\,q\,\Gamma_3 \;,
    \label{eq:jq_1w}
\end{equation} 
The  value of  $q$ can be derived by maximizing the self-consistency relation describing the superconductor with respect to $q$ \cite{baumard_non-uniform_2019}: 
\begin{equation}
    q = 14\,m^2\,\alpha^3\,h_x^2\,h_z\,\frac{\Gamma_7}{\Gamma_3} \;.
    \label{eq:q_1w}
\end{equation} 
After substitution of this value into Eq. (\ref{eq:jq_1w}) one finds that $j_q$ cancels exactly the anomalous current $j_\text{an}$. This confirms that the true ground-state of an infinite superconducting wire is a zero-current state \cite{dimitrova_theory_2007}.

In the case of a finite wire, clearly no charge current can flow. This means that the anomalous current, Eq (\ref{eq:j0 1w}) has to be compensated by a phase gradient described by Eq. (23). In other words such a wire with the properties above  will exhibit  different phase values  at both ends and therefore can be  a realization of a phase-battery\cite{goldobin_josephson_2011,strambini_josephson_2020}.

\section{\label{sec:two wires} The two wire system}
\begin{figure}
    \centering
    \includegraphics[scale=0.4]{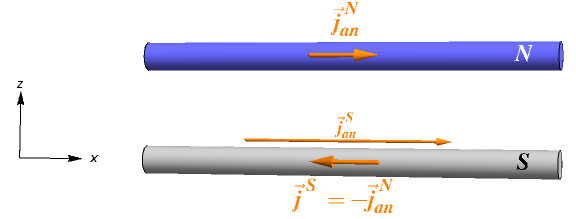}
    \caption{The magnetoelectric effect leads to two currents flowing in the same direction in each of the wires, $j_\text{an}^\text{N}$ and $j_\text{an}^\text{S}$. In the ground state, however the superconducting order parameter has a finite phase gradient $q$, which induces a current $j_q^\text{S}$ in the opposite direction. The total current in S, $j^\text{S} = j_\text{an}^\text{S} + j_q^\text{S}$, then cancels the current $j_\text{an}^\text{N}$ in the normal wire (zero-current state).}
    \label{fig:currents_2w}
\end{figure}

Here we consider a different setup in which superconductivity and spin-dependent fields are spatially separated in two different wires, see Fig. \ref{fig:two_wires}. The coupling of the wires is described via a spin-independent hopping term with amplitude $t$. We assume as before an inhomogeneous order parameter in the superconducting wire and from the self-consistency relation we calculate the phase gradient in the inhomogeneous case superconductivity.  
Separately we compute the anomalous charge current and show that it is exactly compensated by the contribution coming from the modulation of the superconducting order parameter. 
In contrast to the one wire case, the total zero-current state corresponds to two counter-flowing finite currents in each wire (see Fig. \ref{fig:currents_2w}).


For the superconducting wire we assume a spatially oscillating order parameter $\Delta (x) = \Delta_0\,e^{\text{i}\,q\,x}$. The second wire is in the normal state and exhibits Rashba SOC and a local exchange field. We assume that these fields are parallel to each other. 

The Hamiltonian of the system has the same form as Eq. (\ref{eq:Hamiltonian 1w}) enlarged over the wire index space such that:
\begin{align}
    \hat{h}_\text{N} =&  \left(\xi + \frac{q^2}{8\,m}\right)\tau_z + \frac{q\,p}{2\,m} + t\,\tau_z\,\eta_x \notag \\ &+ \left(\alpha\,p\,\tau_z + \frac{\alpha\,q}{2} + h_z\right)\sigma_z\,\frac{\eta_0 + \eta_z}{2}  \;,
    \label{eq:h_N 2w}
\end{align}
where $t$ is the hopping energy and $\eta_{x,y,z}$ are the Pauli matrices  in the  wire index space. To simplify, we assume that the chemical potential is the same in both wires. The BCS Hamiltonian  $\hat{h}_\text{BCS}$ is expressed as:
\begin{equation}
    \hat{h}_\text{BCS} = - \Delta_0\,\tau_x\,\frac{\eta_0 - \eta_z}{2} \;.
    \label{eq:h_BCS 2w}
\end{equation}
As in Sec. \ref{sec:one wire}, we assume that the temperature is closed to the critical temperature $T_c$, such that $\Delta_0\ll T$ and can be treated perturbatively. At second order in $\Delta_0$ the Green's function describing the double wire system is obtained from Eq. (\ref{eq:G 1w}), where $G_\text{N}$ can be written in Nambu-spin basis as
\begin{equation}
    G_\text{N} = \begin{pmatrix}
        G_{--} && 0 && 0 && 0 \\
        0 && G_{-+} && 0 && 0\\
        0 && 0 && G_{+-} && 0\\
        0 && 0 && 0 && G_{++} 
    \end{pmatrix} \;,
    \label{eq:Gn_2w}
\end{equation}
where $G_{\lambda\kappa}$ are 2$\times$2 matrices in the wire space, and  $\lambda = \pm 1$ ($\kappa = \pm 1$) corresponds to the Nambu (spin) indices. Specifically, {the coefficients $G_{\lambda\kappa}$ read} : 
\begin{widetext}
\begin{equation}
    G_{\lambda\kappa} = \frac{\text{i}\,\omega_n + \lambda\left(\xi + \frac{q^2}{8\,m}\right) - \frac{q\,p}{2\,m} + \kappa\left(h_z - \lambda\,\alpha\,p + \frac{\alpha\,q}{2}\right)\frac{\eta_0 - \eta_z}{2} + \lambda\,t\,\eta_x}{\left[\text{i}\,\omega_n + \lambda\left(\xi + \frac{q^2}{8\,m}\right) - \frac{q\,p}{2\,m}\right]\left[\text{i}\,\omega_n + \lambda\left(\xi + \frac{q^2}{8\,m}\right) - \frac{q\,p}{2\,m} + \kappa\left(h_z - \lambda\,\alpha\,p + \frac{\alpha\,q}{2}\right)\right] - t^2} \;.
    \label{eq:Glambda_2w}
\end{equation}
\end{widetext}
The total Green's function is then obtained by substituting Eqs. (\ref{eq:Gn_2w}-\ref{eq:Glambda_2w}) into Eq. (\ref{eq:G 1w}).

We now use the self-consistency relation written in terms of the Green's function $G$ to derive the expression of the wave-vector $q$ near the emergence of the modulated phase: 
\begin{equation}
\Delta_0 = \frac{|\gamma|\,T}{4}\,\displaystyle\sum_{\omega_n} \int_{-\infty}^{+\infty}\frac{\mathrm{d}p}{2\,\pi}\,\text{Tr}\left(\frac{G}{\Delta_0}\,\tau_x\,\frac{\eta_0 - \eta_z}{2}\right)\;,
\label{eq:SCE1}
\end{equation}
where $\gamma$ is the effective attractive electron-electron BCS coupling constant. Close to the critical temperature, Eq. (\ref{eq:SCE1}) can be written in a more convenient form, eliminating $\gamma$ \cite{daumens_inversion_2003,tollis_inversion_2005}:
\begin{align}
&\ln\left(\frac{T_c}{T_{c0}}\right) = 2\,T_c\notag\\ &\times\sum_{\omega_n > 0}\left[\frac{v_\text{F}}{8}\, \text{Re}\left(\int_{-\infty}^{+\infty}  \text{Tr}\left(\frac{G}{\Delta_0}\,\tau_x\,\frac{\eta_0 - \eta_z}{2}\right) \mathrm{d}p \right)- \frac{\pi}{\omega_n}\right]\;,
\label{eq:eq. SC}
\end{align}
where $T_{c0}$ is the critical temperature of the isolated superconducting wire, i.e. for $t = 0$.

We assume that the Zeeman and spin-orbit interactions, as well as tunneling are small such that $t,\, h_z,\, \alpha\,p_\text{F} \ll T_c$, and again that the chemical potential is the largest energy involved: $\mu \gg T_c,\, E_\text{so}$. We then expand the Green's function $G$ in  Eq. (\ref{eq:G 1w}) in series of $h_z$, $\alpha\,p$ and $t$, after replacing Eqs. (\ref{eq:Gn_2w}-\ref{eq:Glambda_2w}). 
Moreover, since the  wave-vector $q$ is small near the emergence of the modulation phase, we also expand $G$ over $q$. 

Only terms even in momentum  survive the momentum integration in Eq. (\ref{eq:eq. SC}). As in the previous section we can then change the integration variable to $\xi$, see Eq. (\ref{eq:change_int}). 
To keep consistently all terms with same power of the small parameter $\xi/\mu$ one has to expand $N\left(\xi\right)$ up to first order. 
Then the self-consistency equation reduces to:
\begin{align}
    \ln\left(\frac{T_c}{T_{c0}}\right) = -2\,\pi\,T_c\,\sum_{\omega_n > 0}&\left[\frac{t^2}{2\,\omega_n^3} + \frac{3}{8\,\omega_n^5}\,\alpha\,h_z\,t^2\,q \right. \notag\\ &\left. + \frac{v_\text{F}^2}{4\,\omega_n^3}\,q^2\right] \;.
    \label{eq:SCEq}
\end{align}
The equilibrium value of $q$ is obtained by maximizing Eq. (\ref{eq:SCEq}) with respect to $q$, which is the equivalent of minimizing the Ginzburg-Landau free energy. We obtain:
\begin{equation}
    q = -\frac{3}{4\,v_\text{F}^2}\,\alpha\,h_z\,t^2\,\frac{\Gamma_5}{\Gamma_3} \;.
    \label{eq:q0 2w}
\end{equation}
Contrary to the single wire system \cite{baumard_non-uniform_2019}, the $x$-component of the field is not needed to generate the inhomogeneous phase. Indeed the two wire system behaves as quasi 2D, as discussed in Sec. \ref{subsec:symmetry 2w}, and therefore one expects to get a magnetoelectric effect also in the case of parallel SOC and exchange field. 

The total charge current in the two-wire system  is given by the sum of the current flowing in the superconducting wire, $j^\text{S} = j_\text{an}^\text{S} + j_q^\text{S}$, and in the normal wire, $j^\text{N} = j_\text{an}^\text{N} + j_q^\text{N}$:
\begin{equation}
    j = j^\text{S} + j^\text{N} \;.
    \label{eq:j 2w}
\end{equation}
In terms of the Green's function $G$, these components read:
\begin{widetext}
\begin{align}
    j^\text{S} &= - \frac{e\,T}{2} \sum_{\omega_n} \int_{-\infty}^{+\infty} \frac{\mathrm{d}p}{2\,\pi}\left[\frac{p}{m} \text{Tr} \left(G\,\frac{\eta_0 - \eta_z}{2}\right) + \frac{q}{2\,m} \text{Tr} \left(G\,\tau_z\,\frac{\eta_0 - \eta_z}{2}\right)\right] \;,
    \label{js}\\
    j^\text{N} &= - \frac{e\,T}{2} \sum_{\omega_n} \int_{-\infty}^{+\infty} \frac{\mathrm{d}p}{2\,\pi}\left[ \frac{p}{m} \text{Tr} \left(G\,\frac{\eta_0 + \eta_z}{2}\right) + \frac{q}{2\,m} \text{Tr} \left(G\,\tau_z\,\frac{\eta_0 + \eta_z}{2}\right) + \alpha \text{Tr} \left(G\,\sigma_z\,\frac{\eta_0 + \eta_z}{2}\right)\right] \;.
    \label{eq:jn}
\end{align}
\end{widetext}
Using Eq. (\ref{eq:change_int}) to compute the integrals over $p$  one obtains the following expressions for the currents:
\begin{align}
    j_\text{an}^\text{S} &=  -e\,T_c\,\Delta_0^2\,\frac{5}{4\,v_\text{F}}\,\alpha\,h_z\,t^2\,\Gamma_5 \;; \label{eq:j_an-s_fin}\\
    j_q^\text{S} &=  -2\,e\,T_c\,\Delta_0^2\,q\,v_\text{F}\,\Gamma_3 \;; \label{eq:jqs_fin}\\
    j_\text{an}^\text{N} &=  -e\,T_c\,\Delta_0^2\,\frac{1}{4\,v_\text{F}}\,\alpha\,h_z\,t^2\,\Gamma_5 \;; \label{eq:j_an-n_fin}\\
    j_q^\text{N} &= 0 \;. \label{eq:jqn_fin}
\end{align}
By replacing $q$ by its expression from  Eq. (\ref{eq:q0 2w}), one can easily check that the total current, Eq. (\ref{eq:j 2w}), is zero.    
However, the current in each wire is finite with $j^\text{S} = -j^\text{N}$, as illustrated in Fig. \ref{fig:currents_2w}.
This is a remarkable result that shows that even though the ground state corresponds to a zero-current state, finite  currents may flow in each of the wires.  

The anomalous current predicted in Sec. \ref{subsec:symmetry 2w} is obtained by imposing  $q = 0$. It is finite and linear in $\alpha$ and $h_z$:
\begin{equation}
    j_\text{an} = j_\text{an}^\text{S} + j_\text{an}^\text{N} =  -\frac{3}{2\,v_\text{F}}\,e\,T_c\,\Delta_0^2\,\alpha\,h_z\,t^2\,\Gamma_5\;.
    \label{eq:jtot_2w}
\end{equation}
This result is in agreement with the symmetry arguments discussed in Sec. \ref{subsec:symmetry 2w}. 

For practical purposes it is convenient to write the last expression  in terms of dimensionless parameters.  After restoring $\hbar$ and $k_B$ we obtain:
\begin{equation}
    j_\text{an} =   -  \gamma \frac{h_z}{\phi_0}\left(\frac{\Delta_0}{k_B T_c}\right)^2\frac{\alpha}{v_F}\left(\frac{t}{k_B T_c}\right)^2\;,
    \label{eq:jtot_2w2}
\end{equation}
where $\phi_0=\pi \hbar/e$ and $\gamma$ is a numerical pre-factor of the order of 0.015. 

\section{\label{sec:conclusion}Conclusion}
We have shown that the interplay between superconductivity, spin-orbit interaction and Zeeman field in 1D systems allows for the existence of spontaneous charge currents. We have first investigated these currents within the SU(2) covariant formulation, and determined the leading order contributions to the anomalous currents in the SOC and Zeeman field, in terms of the magnetic and electric SU(2) fields.  
In the case of a single wire we confirmed that a finite anomalous current can only appear when the Zeeman field has a component parallel and one orthogonal to the SOC \cite{ojanen_magnetoelectric_2012, nesterov_anomalous_2016}. In the case of a two wire system, where the superconducting and spin-dependent correlations are spatially separated, there is a finite SU(2) magnetic field and we predict that a Zeeman field parallel to the spin-orbit field is sufficient to induce an anomalous current.   

We confirmed these results by computing explicitly the anomalous currents in the framework of Green's function formalism for both setups. Moreover, allowing an inhomogeneous superconducting order parameter, we found that the ground state corresponds, in fact, to a total zero current state where the anomalous contribution is compensated by the contribution from the modulated superconducting order parameter (characterized by a finite wave vector). In the two wire system, however, the total zero-current state consists of two opposite currents flowing in each of the wires.

Experimentally, ballistic double wire setups have been realized in hybrid semiconducting systems \cite{auslaender_spin-charge_2005,auslaender_tunneling_2002, steinberg_charge_2008, scheller_possible_2014}.  Such systems could be proposed as a good platform to realize the two wire system described in Sec. \ref{sec:two wires}. One could imagine to induce superconductivity in one of the wires by proximity effect from an adjacent superconductor. 
{  The study of such a setup  can be an  interesting extension of our work. In that case, superconducting correlations are  imposed by the bulk superconductor and therefore  no self-consistency relation is needed.  We still expect a zero-current ground state with opposite currents flowing in each wire. The exact distribution of the currents in such system  will depend on the exact geometry } In addition, because the existence of the magnetoelectric effect is independent of the microscopic details of the systems and can be explained by symmetry arguments only (see  section \ref{sec:symmetry}). In this regard we expect that magnetoelectric effects  can be observed in two metallic wires, for example a conventional superconductor as Al, and a heavy metal with SOC, as Pt.  If the wires are tunnel coupled to each other one  expects opposite supercurrents flowing in each wire, as described in section \ref{sec:two wires}.
In order to estimate the magnitude of these currents in a  small metallic wire we multiply the contribution of  a single conducting channel, Eq. (\ref{eq:jtot_2w2}), by the number of transverse modes $\sim {\cal S}k_F^2$, where ${\cal S}$ is the cross-section of the wire.  If we assume  $\Delta_0/T_c=1$, $\alpha/v_F=0.1$ ,  $t/T_c=0.5$,  $h_z=0.1$meV,  $k_F=2. 10^{10}$m$^{-1}$, and ${\cal S}=30-300$nm$^2$, we estimate currents of the order of  36-360  nA.  This value can be significantly enhanced by going beyond our approximation of large temperatures and choosing materials with large spin-orbit coupling.   If the wire is embedded in a superconducting loop the anomalous current will manifest as a sizeable   phase  shift in the phase-current relation as reported recently in  several experiments \cite{szombati_josephson_2016,murani_ballistic_2017,assouline_spin-orbit_2019,strambini_josephson_2020}.

\section*{Acknowledgements}
We thank Ilya Tokatly for interesting  discussions. J. B. and F. S. B. acknowledge funding by the Spanish Ministerio de Ciencia, Innovaci\'on y Universidades (MICINN) under Projects No. FIS2014-55987-P and No. FIS2017-82804-P. J. B. acknowledges the financial support from the Initiative d'Excellence (IDEX) of the Universit\'e de Bordeaux. This work was supported by EU Network COST CA16218 (NANOCOHYBRI) and the French ANR projects SUPERTRONICS and OPTOFLUXONICS (A. B. and J. C.).

\bibstyle{apsrev4-1}
\bibliography{References}

\end{document}